\documentstyle[epsfig]{elsart}

\begin{document}

\begin{frontmatter}
\author{I. I. Mazin$^{a,b}$ and D. J. Singh$^a$}
\address{$^{a}$Complex Systems Theory Branch, Naval Research Laboratory, Washington,\\
DC 20375-5320\\
$^{b}$Computational Science and Informatics, George Mason University,\\
Fairfax, VA}
\title{Magnetism and spin-fluctuation induced superconductivity in ruthenates.}
\centerline{Talk presented at the SNS'97 Conference, Cape Cod, 1997}
\begin{abstract}
Layered and pseudocubic Ru-based perovskites have been the subject of
considerable recent
attention, due to their unusual magnetic
properties and the discovery of superconductivity in one member of the family, Sr%
$_{2}$RuO$_{4}$. From a magnetic point of view,  interest derives from the
stable ferromagnetism in SrRuO$_{3}$, gradually disappearing to a
non-magnetic phase upon substituting Sr with isovalent Ca, a very unusual 
 kind of behavior for 3d perovskites. On the superconducting side, 
interest was stimulated by theoretical conjectures and experimental
indications that Sr$_{2}$RuO$_{4}$ might be a $p$-wave superconductor. We
report first-principles LSDA calculations for ferromagnetic SrRuO$_{3}$,
antiferromagnetic Sr$_{2}$YRuO$_{6}$, non-magnetic CaRuO$_{3}$, and
superconducting Sr$_{2}$RuO$_{4}$. In all cases, magnetic properties are well
reproduced by the calculations. Anomalous properties are explained
in terms of simple TB models and Stoner theory. An important result 
 is that O bears sizable magnetic moments and plays an important
role in the formation of the magnetic states. Based on these
calculations, we have built a model for the $q$-dependent Stoner interaction,
which we consequently applied to Sr$_{2}$RuO$_{4}$   to estimate
superconducting and mass-renormalization electron-paramagnon coupling
constants. We found that spin-fluctuation induced $p$-wave
superconductivity is possible in Sr$_{2}$RuO$_{4}$. The estimated critical
temperature, specific heat and susceptibility renormalizations are all in
good agreement with  experiment.
\end{abstract}
\end{frontmatter}
The recent discovery of superconductivity in the layered ruthenate,
Sr$_{2}$RuO$_{4}$ \cite{chi} has generated new interest in
 Ru-based perovskites. At 
first glance this material seems  analogous to the high-$T_{c}$ cuprates. For 
instance, it has a similar crystal
structure (it is isostructural with La$_{2}$CuO$_{4})$ and is apparently
close to a magnetic instability (Sr$_{x}$Ca$_{1-x}$RuO$_{3}$ and Sr$_{2}$RuYO%
$_{6}$ are ferro- and antiferromagnetic, respectively).
On the other hand, the more we learn
about ruthenates, the less similar to cuprates they seem. While initial
interest was largely related to the similarity to the
high-$T_{c}$ materials, now it is more
that ruthenates are deemed interesting {\it per se}, and, at least in their
magnetic properties they are more variegated and probably more interesting
than cuprates. Furthermore, it appears that superconductivity in Sr$_{2}$RuO$%
_{4}$ can hardly be understood without a good understanding of magnetism in
ruthenium perovskites in general. Thus this paper naturally breaks into two
parts. First, we discuss magnetism in ruthenates, specifically
antiferromagnetic Sr$_{2}$RuYO$_{6}$, ferromagnetic SrRuO$_{3},$ and
paramagnetic CaRuO$_{3}.$ We will show that despite the wide range
of magnetic properties, they all are governed by a simple Stoner-type
mechanism, which manifests itself differently depending on crystal structure.
We then shall show how closeness to a ferromagnetic instability can produce
a triplet superconductivity in Sr$_{2}$RuO$_{4}$ and explain its
normal-state transport properties. We shall also discuss what is currently
maybe the most intriguing question in the theory of superconductivity in Sr$%
_{2}$RuO$_{4},$ namely why the experiment shows finite electronic 
density of states at zero
energy (in NMR and specific heat experiments) at as low as 0.3$T_{c}.$

\subsection{Magnetism}

The great majority of magnetic transition metal 
oxides are based on the 3$d$-series.
Density functional theory in its standard local spin density approximation
(LSDA) does not work very well
for some of these materials; it often fails to yield
the correct magnetic ground state, in many cases it underestimates the magnetic
moments, in some others it does not reproduce correct insulating
behavior. In such cases it  is customary to speak about
 ``strong correlation behavior''. The LSDA
is essentially a mean field theory where electron-electron interactions are
treated in an averaged way, and the nature of a magnetic instability is 
related to the standard Stoner model, where the paramagnetic susceptibility,
renormalized in the RPA-like manner, may diverge at some wave vector. On the
other hand, in the strong correlation picture the zero order
approximation is the large-$U$
Hubbard Hamiltonian with an inherent antiferromagnetic instability  to it 
{\it via} the superexchange mechanism. The first thing to decide is which of
the two basic approaches serves better as the starting approximation.

An important mechanism for magnetic instabilities in  a one-electron framework
is the ``Stoner model''. This is a purely itinerant magnetism approach.
In the LSDA the total energy is written as $%
E=T_{s}+E_{H}+E_{e-i}+E_{xc},$ where $T_{s}$ is the single-particle 
kinetic energy, $E_{H},$ $E_{e-i},$ and $E_{xc}$ are 
the Hartree, the electron-ion, and the exchange-correlation energies,
respectively. A ferromagnetic instability is, in this model,
 an instability with
respect to a perturbation consisting of splitting the band by an exchange
field, readjusting the Fermi level, and recalculating of $E_{xc}$ taking
into account the created magnetic polarization. It is easy to see that the
energy between  ferromagnetic and the paramagnetic states in the lowest
order in magnetization $M$ is 
\begin{equation}
\Delta E=\frac{M^{2}}{4N(0)}-\frac{M^{2}}{4}\frac{\delta ^{2}E_{xc}}{\delta
m^{2}}.  \label{fStoner}
\end{equation}
The last variation, $I=\delta ^{2}E_{xc}/\delta m^{2},$ is called the 
Stoner parameter. It defines the renormalization of the paramagnetic
susceptibility due to spin fluctuations, $\chi =$ $\chi _{0}/(1-I\chi _{0}).$
Note that when the exchange splitting is imposed upon a compound with more than
one component, the total magnetization is expressed as $M=\sum_{i}M_{i},$
where $M_{i}$ is the magnetic moment of the $i$-th component and is proportional to
its partial DOS at the Fermi level, $M_{i}/M=N_{i}(0)/N(0).$ This lets one
 relate the average Stoner factor for a compound, $I$, with the Stoner
factors of the constituent atoms: $4\Delta
E_{xc}=-\sum_{i}M_{i}^{2}I_{i}=-M\sum_{i}(N_{i}/N)^{2}I_{i},$ hence $%
I=-\sum_{i}(N_{i}/N)^{2}I_{i}.$ Of course, actual LSDA calculations 
take into account distortions of the bands as a function of magnetization,
as well as the higher order in $M$ terms, neglected in the Stoner model.

Looking at such ruthenates as Sr$_{2}$RuYO$_{6}$, SrRuO$_{3},$ and CaRuO$%
_{3} $ from the Stoner point of view, one observes that oxygen $p$-character
is present at the Fermi level to a substantially greater extent
 than in the cuprates or most 3$d$ oxides.  Calculating
the average $I$ for these compounds one finds that the oxygen contribution, $%
[N_{{\rm O}}(0)/N(0)]^{2}I_{{\rm O}}$ is substantial; if it is neglected,
the Stoner criterion $IN(0)\geq 1$ is not satisfied for any of them.
If it is included, Sr$_{2}$RuYO$_{6}$ and SrRuO$_{3}$ appear to
be unstable against ferromagnetic transitions, while CaRuO$_{3},$ because of
a slightly different DOS, is barely stable. Detailed analysis of the magnetism
in these compounds has been published elsewhere \cite{we}. The key
ingredient is the strong Ru-O hybridization, which puts O character at $E_F$
and assures the validity of the Stoner model.

One can   generalize Stoner approach to antiferromagnetic
instabilities. The main difference from the ferromagnetic case is that the DOS
in the Stoner formula has to be replaced by the one-electron
susceptibility, $N(0)=\chi (0)\rightarrow \chi (Q),$ where $Q$ is the
antiferromagnetic vector. What turns out to be important is that if the AFM
ordering in question is such that some atoms do not bear a magnetic moment by
symmetry, they should be excluded from the calculation of the average Stoner
factor. This is the case in SrRuO$_{3},$ and CaRuO$_{3}$ where oxygen,
bridging two nearest neighbor Ru, cannot acquire a magnetic moment if the two
Ru atoms are aligned antiferromagnetically. Correspondingly, the average $I$
for antiferromagnetically ordered
 (Sr,Ca)RuO$_{3}$ would be considerably smaller than for
ferromagnetic analogues. In Sr$_{2}$RuYO$_{6}$ there are no bridging oxygens
and the ground state is antiferromagnetic, with the oxygens bearing a large
fraction of the total magnetization. This is reproduced
by detailed self consistent LSDA calculations.

\subsection{Superconductivity}

LSDA calculations for  Ru-based perovskites generally either
 predict a magnetic ground
state or a paramagnetic state very close to an instability. The quasi-2D Sr$%
_{2}$RuO$_{4}$ is not an exception --- LSDA calculations give a Stoner
renormalization $(1-NI)^{-1}=9$ (experiment gives similar numbers). Thus,
one expects strong spin fluctuations to be present in this compound. The
situation is similar to Pd metal, where $NI$ is also close to 1. It is very
hard to expect that a conventional superconducting state would survive in the
presence of such spin fluctuations. In fact, Pd has a sizable electron-phonon
interaction and would have been a superconductor apart from spin fluctuations,
and in fact becomes such in
amorphous state where spin fluctuations are suppressed\cite{bose}. On the
other hand, it is known (see, e.g., Ref.\cite{legett}) that spin fluctuations
provide effective repulsion for the singlet ($s,$ $d$) pairing, but
attraction for triplet ($p$) pairing. Thus it is tempting to ascribe
superconductivity in Sr$_{2}$RuO$_{4}$ to the spin-fluctuation induced $p$%
-wave pairing\cite{rice,jap}. LSDA calculations can be used as a tool to get a
feeling about the size of the attraction provided by exchange of spin
fluctuations and whether it is sufficient to explain the superconducting and
normal state properties of this material.

The valence bands of Sr$_{2}$RuO$_{4}$ are formed by the three $t_{2g}$ Ru
orbitals, $xy,$ $yz,$ and $zx.$ These are hybridized with the in-plane
oxygen and, to a considerably
lesser extent, with the apical oxygen\cite{singh,oguchi} $p$%
-states. The bare oxygen $p$ levels are well ($\sim 2$ eV) removed from $%
E_{F}$, so the effect of the O $p$ orbital is chiefly renormalization of the
Ru $t_{2g}$ levels, and assisting in the $d-d$ hopping. With nearest
neighbors only, this gives one nearly circular cylindrical electronic sheet (%
$\gamma )$ of the Fermi surface (FS) and four crossing planes (quasi-1D FS).
The weak $xz-yz$ hybridization reconnects these planes to form two
tetragonal prisms, a hole one ($\alpha )$ and an electron one ($\beta )$. De
Haas-van Alphen experiments confirm this fermiology\cite{dHvA}. In fact, the
LDA $\alpha ,$ $\beta ,$ and $\gamma $ areas deviate from the
dHvA experiment by only -2\%, -3\% and 5\% of the Brillouin zone area,
respectively, and an exact match can be achieved by very slight shifts of
the bands $\alpha ,$ $\beta ,$ and $\gamma $ by 5, -4, and -3 mRy,
respectively. Such agreement is generally considered very good even in
simple metals, and the small mismatch (which does not change the FS
topology) is may be due to some underestimation in LDA calculations of the
tiny $xz-yz$ hybridization. Both calculation and experiment give nearly two
dimensional Fermi surface: the relative $c$-axis variation
 of the extremal cross-section
areas of the sheets $\beta $ and $\gamma $ is 6\% and 1.5\%, respectively
(for these two sheets the extremal cross-sections are in the planes $k_{z}=0$
and $k_{z}=\pi /c).$ For the sheet $\alpha $ the relative change is 2\% (for
this sheet the extremal cross-sections are in the planes $k_{z}=0$ and $%
k_{z}=\pi /2c).$ Experiment gives the numbers about twice smaller for all
three sheets\cite{andy-unp}; the difference is larger than the computational
error, and presumably has its origin in the effects beyond the density
functional theory. We repeated the calculations using two non-LDA techniques,
generalized gradient approximation\cite{GGA} and weighted density
approximation\cite{WDA}, but the numbers hardly changed. In the following
all calculational results are from the LDA linearized augmented
plane wave calculations%
\cite{singh}.

We assume that the exchange of the spin fluctuations is
responsible for superconductivity (and for the mass renormalization, to be
discussed later). Such an interaction in metals was studied with respect to
possible superconductivity in Pd in the late 1970-ties (see, e.g.,\cite
{appel,AM}), and later in connection with heavy fermions. Assuming the
Migdal theorem (a common
approximation, although not well justified for spin fluctuations), the
parallel-spin interaction, relevant for triplet pairing is given by the sum
of the bubble diagrams with odd numbers of loops, 
\begin{equation}
V({\bf q=k-k}^{\prime })=\frac{I^{2}(q)\chi _{0}(q)}{1-I^{2}(q)\chi
_{0}^{2}(q)}.  \label{V}
\end{equation}

Here $\chi _{0}$ is the one-electron susceptibility, given as 
\begin{equation}
\chi _{0}({\bf q})=\sum_{{\bf k}\alpha \beta }\frac{f_{{\bf k}\alpha }-f_{%
{\bf k+q},\beta }}{\epsilon _{{\bf k}\alpha }-\epsilon _{{\bf k+q,}\beta }}%
\left\langle {\bf k}\alpha |\exp (i{\bf qr})|{\bf k+q,}\beta \right\rangle
^{2},
\end{equation}
with the usual notations. We used the approximation\cite{wePRL} $%
\chi _{0}({\bf q})=\chi _{0}(0)=N(0);$ this is a good approximation for an
isotropic two-dimensional Fermi liquid\cite{chi2D}; we are currently 
investigating the quality of this approximation for Sr$_2$RuO$_4$,
which is a two-dimensional, but not isotropic, Fermi liquid (so some
modification of $\chi
(q)$ due to Fermi surface nesting may be expected). In any
case, the $q$-dependence of $I(q)$ is to be taken into account.
 As discussed in the
previous section, for the antiferromagnetic arrangement $I_{AFM}\equiv I(\pi
/a,\pi /a)=I_{Ru}\left( N_{Ru}/N\right) ^{2},$ while $I_{FM}\equiv
I(0)=I_{Ru}\left( N_{Ru}/N\right) ^{2}+2I_{O}\left( N_{O}/2N\right) ^{2}.$
Atomic Stoner factors for Ru and O ions are calculated in a standard way and
are $I_{Ru}\approx 0.7$ eV, $I_{O}\approx 1.6$ eV. We found $I_{AFM}$ to be
smaller than $I_{FM}$ by 14\% (oxygen contribution $\Delta I=0.06$ eV). A $q$%
-dependence that reflects this effect is $I(q)=I/(1+b^{2}q^{2}),$ where $%
b^{2}=0.5(a/\pi )^{2}\Delta I/(I-\Delta I)\approx 0.08(a/\pi )^{2}.$

Using these numbers, we calculate the effective coupling
constant in $p$-channel. Following the suggestion of Agterberg {\it et al}%
\cite{Agt}, we calculate the coupling constants separately for the
three bands in question: $xy$ ($\gamma ),$ $yz$ ($\zeta ),$ and $zx$ ($\xi ).
$ The corresponding formula is 
\begin{equation}
\Lambda _{ij}^{p}=(N_{i}N_{j}/N)\langle V({\bf k-k}^{\prime })({\bf v}_{{\bf %
k}}^{i}\cdot {\bf v}_{{\bf k}^{\prime }}^{j})/(v_{{\bf k}}^{i}v_{{\bf k}%
^{\prime }}^{j})\rangle _{ij},  \label{lamp}
\end{equation}
where $i$ and $j$ label the three bands, and ${\bf v}$ is the Fermi
velocity. By symmetry, the coupling matrix is
\begin{equation}
\left( 
\begin{array}{ccc}
\Lambda _{\gamma \gamma }^{p} & \Lambda _{\gamma \xi }^{p} & \Lambda
_{\gamma \xi }^{p} \\ 
\Lambda _{\gamma \xi }^{p} & \Lambda _{\xi \xi }^{p} & 0 \\ 
\Lambda _{\gamma \xi }^{p} & 0 & \Lambda _{\xi \xi }^{p}
\end{array}
\right) ,  \label{lam}
\end{equation}
and we calculate $\Lambda _{\gamma \gamma }^{p}=0.16,$ $\Lambda _{\xi \xi
}^{p}=0.075,$ and $\Lambda _{\gamma \xi }^{p}=0.025.$ The critical
temperature is defined by the maximum eigenvalue of the matrix ($%
N/N_{i})\Lambda _{ij}^{p}$\cite{allenB}. The corresponding eigenvector
defines the relative magnitude of the order parameter in bands $\gamma $ and
($\xi ,\zeta )$ near $T_c$. We find the maximum eigenvalue of the corresponding
coupling matrix is $\lambda _{p}=$0.43, and the corresponding
superconducting state is $0.85\gamma +0.38\xi +0.38\zeta .$ It is worth
noting that using notations of Ref.\cite{Agt}, and taking into account the
partial DOS $N_{\gamma }:N_{\xi }:N_{\zeta }=0.44:0.28:0.28,$ the matrix (%
\ref{lam}) can be translated to the interaction matrix $U$ as 
\begin{equation}
{\bf U}=\left( 
\begin{array}{ccc}
u_{\gamma \gamma } & u_{\gamma \alpha } & u_{\gamma \beta } \\ 
\ u_{\gamma \alpha } & u_{\alpha \alpha } & u_{\alpha \beta } \\ 
\ u_{\gamma \beta } & u_{\alpha \beta } & \ u_{\beta \beta }
\end{array}
\right) ,
\end{equation}
where $u_{\gamma \gamma }:u_{\gamma \alpha }:u_{\gamma \beta }:u_{\alpha
\alpha }:u_{\alpha \beta }:u_{\beta \beta }=0.96:0.08:0.16:0.25:0.51:1,$ to
be compared with the value conjectured in Ref.\cite{Agt}, $%
0.09:0.09:0.09:1:1:1.$ Their hypothesis about the smallness of the nondiagonal
elements $u_{\gamma \alpha }$ and $u_{\gamma \beta }$ is confirmed by the
calculations, but the assumption about the smallness of $u_{\gamma \gamma }$
is not. In any event, the calculated value of $\lambda _{p}=$0.43 is
sizable, and sufficient to explain the observed superconductivity. We would
like to emphasize the role of oxygen in this scenario: if not for the oxygen
Stoner factor, the $q$-dependence of the effective interaction $V(q)$ would
be so small that the coupling in Eq. (\ref{lamp}) would average near zero.

\subsection{Renormalization}

The mass renormalization is not as easy to define. Besides the parallel-spin
interaction (\ref{V}), there is the antiparallel-spin interaction, given in
the same approximation by the sum of the chain diagrams with even numbers of
loops, plus ladder diagrams \cite{appel,BE}. In the case of a 
contact interaction,
the total interaction is three times stronger than the interaction in the
parallel-spin channel only. It was pointed out\cite{AM}, though, that there
is no good physical reason to single out any particular class of diagrams.
It was found that including all three classes above leads to systematic
overestimation of mass renormalizations by a factor of 2 to 3 \cite
{appel,Levin}. The present case is further complicated because unlike the
electron-phonon interaction, the electron-electron (and, correspondingly,
the electron-paramagnon) interaction is already included in some average way
in the LSDA band structure. Thus, the electron-paramagnon mass
renormalization is to some extent included in the LDA mass as well.

Despite all these difficulties, one can get an idea about the size of the
electron-paramagnon mass renormalization by making calculations with the
parallel-spin interaction (\ref{V}) only. The mass
renormalization then is computed in the same way as the electron-phonon
renormalization, {\it i.e}, by taking the average of $V({\bf q)}$ of Eq.(\ref
{V}) over the FS. One has to remember, though, that there are other effects
beyond the LDA, apart from the one that we calculate, which may further increase
the observable mass.

The coupling matrix which defines mass renormalization is  written as $%
\Lambda _{ij}^{s}=(N_{i}N_{j}/N)<V({\bf k-k}^{\prime })>_{ij},$ and the mass
renormalization in band $i$ is defined as $\lambda _{i}^{s}=\nu
_{i}^{-1}\sum_{j}\Lambda _{ij}.$ The average mass renormalization is $%
\lambda ^{s}=$ $\sum_{ij}\Lambda _{ij}^{s}.$ We calculate $\Lambda _{\gamma
\gamma }^{s}=0.35,$ $\Lambda _{\xi \xi }^{s}=0.32,$ $\Lambda _{\gamma \xi
}^{s}=0.16,$ $\Lambda _{\xi \zeta }^{s}=0.03.$ This gives $\lambda _{\gamma
}^{s}=(\Lambda _{\gamma \gamma }^{s}+2\Lambda _{\gamma \xi }^{s})/\nu
_{\gamma }=1.5,$ $\lambda _{\xi }^{s}=(\Lambda _{\xi \xi }^{s}+\Lambda
_{\gamma \xi }^{s}+\Lambda _{\xi \zeta }^{s})/\nu _{\xi }=1.8,$ $\lambda
^{s}=1.7,$ to be compared with experimental dHvA values of 3, 2.3, and 3,
respectively. The difference may be due to an electron-phonon coupling of
the order of 1 and/or antiparallel spin fluctuations, neglected in our
calculations, as well as to the omission of the non-Migdal diagrams. In view of
the  underlying approximations, the agreement is fairly good.

One of the key problems, as discussed in Refs. \cite{sigrist,Agt}, is the
residual electronic specific heat\cite{sheat}, which remains at about 50\%
of its normal value well into the superconducting regime. There are
superconducting solutions (``nonunitary states'') for triplet pairing
that are gapless, that is, have finite density of states at zero energy and
zero temperature. However, the pairing energy for such states is lower than
for the gapped states considered above. This led Agterberg {\it et al} \cite
{Agt} to postulate a pairing matrix that yields a vanishing gap for the $%
\gamma $ band. This, however, does not square with the quantitative estimate
presented here. An earlier assumption\cite{sigrist,jap} was that
the excess pairing energy that forbids nonunitary combination of the order
parameters may be overcome by additional magnetic (Stoner) energy
in a nonunitary state. The requirements are strong Stoner renormalization
(supported by the calculations) and strong particle hole asymmetry\cite{ferro}%
. However, a quantitative estimate according to Ref.\cite{ferro} shows that
the effect is by far too weak. The criterion is $\left[ \frac{T_{c}d\log N}{%
dE_{F}}\right] ^{2}\frac{1}{1-IN}\log \frac{\omega _{sf}}{T_{c}}\sim 10^{-5},
$ while it should be of the order 1 for the nonunitary state to exist.

Another possibility is related to an observation made a decade ago in
connection with the high-$T_c$ superconductivity\cite{GD}: A well-known
fact is that virtual phonons, even in a strongly coupled system, have no
pair-breaking effect, so that the density of states remains zero below
the gap at zero temperature in a clean superconductor. However, this is a 
consequence of an internal symmetry of the Eliashberg equations, namely that
the coupling function $\alpha^2 F(\omega)$, entering the equation on $\Delta$,
is the same as  $\alpha^2 F(\omega)$, entering the equation on $Z$. In case
of $p$-wave pairing, for instance, this is not true any more, and formally
there is finite density of states inside the gap at any temperature. 
Unfortunately, direct calculations\cite{Golun} show that this effect is
quantitatively strong only if a noticeable part of  $\alpha^2 F(\omega)$
exists at $\omega < \Delta$, which is not the case in Sr2RuO4.

Maybe the simplest explanation of the ``residual DOS mystery'' is still 
the most plausible. Despite the large mean free path, which in the reported
1.35 K 
samples reaches 1500-2000 \AA \cite{andy},
this superconductor is still in the dirty
limit: the Abrikosov-Gor'kov pair-breaking parameter $\gamma=1/2\tau
\Delta=\pi \xi_0/2 l_{\rm m.f.p.}=0.7$, using the value for the coherence 
length $\xi_0=1000$ \AA \cite{maeno}. Nonmagnetic impurities in a unitary
2D $p$-wave superconductor act as magnetic impurities in an $s$-wave
superconductor. The DOS is given by the standard expression
\[
N(E)/N_{\rm norm}=Re{u(E)\over\sqrt{u(E)^2-1}}
\]
where $u(E)$ satisfies the equation
\[E=u-\gamma x/\sqrt{1-x^2}\]

The resulting DOS at $T=T_c/3$ is shown on Fig.1 and is seen to be very
large below the gap (and does not show any trace of piling of the DOS
above the gap). 
\subsection{Conclusions}

To summarize, we have presented first principles calculations indicating
that interactions due to exchange of FM spin fluctuations, as calculated
from the LDA band structure, are sufficiently strong to explain both the
mass renormalization and superconducting critical temperature of Sr$
_2$RuO$_4$.

This work was supported by the ONR. Computations were performed at the DoD
HPCMO NAVO and ASC facilities.

\begin{figure}[tbp]
\vskip 2mm
\caption{Relative density of states at $T=0.3 T_c$ in Abrikosov-Gor'kov
theory for pair-breaking parameters $\gamma=$0, 0.07, and 0.7. We estimate
that for 1.35 K superconducting samples the pair breaking parameter
$\gamma$ is at least 0.7.}
\label{figchi}
\end{figure}

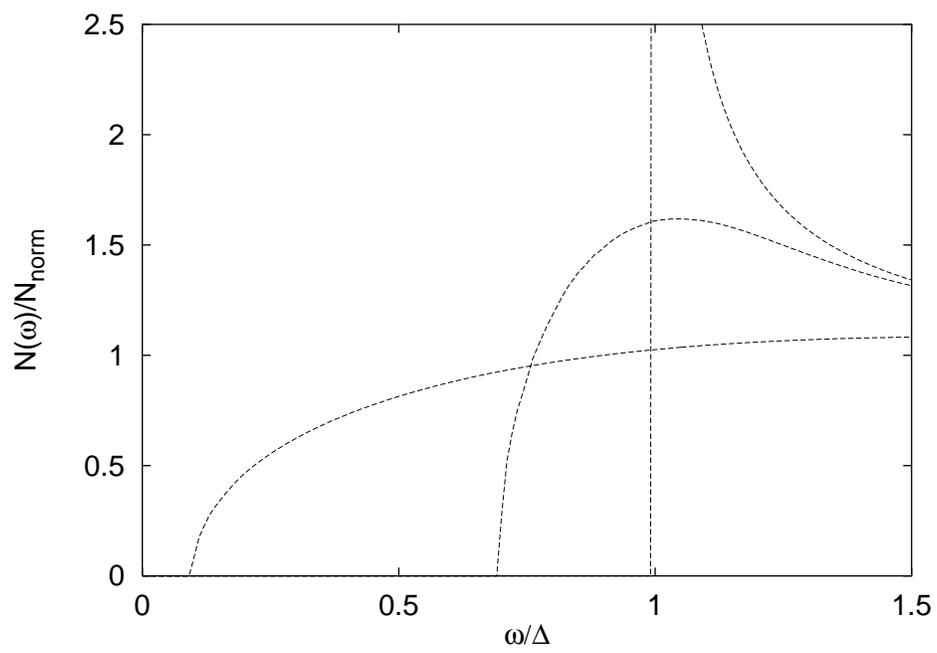
\end{document}